# Exoplanet atmospheres with EChO: spectral retrievals using EChOSim

Joanna K. Barstow[1,*], Neil E. Bowles[1], Suzanne Aigrain[1], Leigh N. Fletcher[1], Patrick G. J. Irwin[1], Ryan Varley[2], Enzo Pascale[3]

[1] Department of Physics, University of Oxford, Oxford, UK
[2] Department of Physics, University College London, London, UK
[3] School of Physics and Astronomy, University of Cardiff, Cardiff, UK
[*] jo.barstow@astro.ox.ac.uk

We demonstrate the effectiveness of the Exoplanet Characterisation Observatory mission concept for constraining the atmospheric properties of hot and warm gas giants and super Earths. Synthetic primary and secondary transit spectra for a range of planets are passed through EChOSim (Waldmann & Pascale 2014) to obtain the expected level of noise for different observational scenarios; these are then used as inputs for the NEMESIS atmospheric retrieval code and the retrieved atmospheric properties (temperature structure, composition and cloud properties) compared with the known input values, following the method of Barstow et al. (2013a). To correctly retrieve the temperature structure and composition of the atmosphere to within 2 $\sigma$, we find that we require: a single transit or eclipse of a hot Jupiter orbiting a sun-like (G2) star at 35 pc to constrain the terminator and dayside atmospheres; 20 transits or eclipses of a warm Jupiter orbiting a similar star; 10 transits/eclipses of a hot Neptune orbiting an M dwarf at 6 pc; and 30 transits or eclipses of a GJ1214b-like planet.

## 1. Introduction

The Exoplanet Characterisation Observatory (EChO; Tinetti et al. 2012) was a proposed European Space Agency mission for the Cosmic Visions M3 launch. Its goal was to survey the atmospheres of a range of extrasolar planets from the L2 Lagrange Point, over a period of 5 years. This would enable us to better understand the thermochemistry, dynamics and evolution of planetary atmospheres. By surveying a range of planets under different environmental conditions, this mission would help us to better understand the climatology, thermochemistry, dynamics and cloud formation processes at work in the ensemble of planetary types. A crucial requirement of the mission is that the temperature structure and composition of a range of planets can be constrained from their visible and infrared primary and secondary transit spectra.

The technique of transit spectroscopy was suggested by Coustenis et al. (1997) and the first detection of a planetary atmosphere in transit was reported by Charbonneau et al. (2002). When an exoplanet passes in front of its parent star as seen from the Earth (primary transit), absorbers within its atmosphere attenuate the starlight, making the planet appear larger in wavelengths where the atmosphere is absorbing. In secondary transit, the planet is eclipsed by the star and the difference between the in- and out-of-transit fluxes allows the thermal and reflected fluxes from the planet to be determined. In principle, these observations allow the absorbers within the planet's atmosphere to be identified. The size of absorption features and the amount of flux from the planet in the infrared also allow constraint to be placed on the thermal structure of the planet's atmosphere, although there are often degeneracies between temperature structure, composition and cloud properties.

In order to determine the number of individual transits/eclipses required for a successful simultaneous retrieval of these properties, it is necessary to calculate the expected level of noise on the spectrum for the different planet cases. This is done using the EChOSim instrument simulation package (Waldmann & Pascale 2014). The package calculates the noise per spectral element for a specified number of transits of each planet at a particular resolution. Three different observing strategies are considered with different spectral resolving powers for wavelengths shorter than 5 μm; *Chemical Census* observations have R=50, *Origin* R=100 and *Rosetta Stone* R=300. These are discussed in more detail in Section 2.1.

We use the Non-linear optimal Estimator for MultivariatE analysis (NEMESIS; Irwin et al. 2008) to generate a series of synthetic spectra for a range of planets. NEMESIS is a radiative transfer model and spectral retrieval tool using optimal estimation (Rodgers 2000), originally developed for solar system planets and recently extended for analysis of exoplanet spectra (e.g. Lee et al. 2012, Barstow et al. 2013a). NEMESIS calculates the expected transit or eclipse spectrum from a given model atmospheric state; it then iterates to the atmospheric state for which the synthetic spectrum provides the best match to the data. In this case, our 'data' are the calculated synthetic spectra; we use EChOSim to calculate the expected noise levels, and we add this noise to the spectra using the method described in Barstow et al. (2013a). We then perform a similar retrieved-versus-input analysis to Barstow et al. (2013a), and from this calculate the number of observations required for each planet for a successful retrieval of the atmospheric state. We require that gas volume mixing ratios are correctly retrieved to within 2 σ, and that the temperature structure is correctly retrieved to within σ at altitudes where there is sensitivity.

## 2. Modelling and retrieval

We use the same set of model planets as Barstow et al. (2013a), with the addition of the warm super-Earth GJ 1214b (Charbonneau et al. 2009), modelled as in Barstow et al. (2013b). These are listed in Table 1. We perform retrievals of spectra corresponding to the three different observation modes of EChO, as outlined in Varley et al. (2014), to investigate how many transits and what spectral resolving power would be sufficient to accurately retrieve the atmospheric state.

We assume all atmospheres have a bulk composition of $H_2$ and He, with trace amounts of $H_2O$, $CO_2$, CO, $CH_4$ and $NH_3$ (except for GJ 1214b, for which we do not include CO or $NH_3$). All line and collision-induced absorption data are as in Barstow et al. (2013a) and references therein. We have modelled GJ 1214b with a H2-He dominated, cloudy atmosphere, and the cloud is assumed to be made of photochemical tholin haze. The cloud particle size distribution and other properties are as in Barstow et al. (2013b).

| Planet | Mass ($10^{24}$ kg) | Radius (km) | $T_{eff}$ (K) | Distance (pc) | Star |
|---|---|---|---|---|---|
| Hot Jup. | 1800 | 75000 | 1500 | 35 | Sun |
| Warm Jup. | 1800 | 75000 | 650 | 35 | Sun |
| Hot Nep. | 180 | 30000 | 800 | 6 | M5 |
| GJ 1214b | 38 | 15350 | 450 | 13 | GJ 1214 |

**Table 1**. Bulk properties, temperature, distance and assumed stellar type for each model planet

We use the NEMESIS spectral retrieval code to perform the retrieval tests. We generate one synthetic primary and one secondary transit spectrum for each planet; these are then fed into EChOSim and the noise level computed for each spectral bin. The noise is then added to the spectrum, assuming it is Gaussian-distributed and uncorrelated. We then use NEMESIS to retrieve the atmospheric state from the noisy spectrum, for both primary and secondary transit. In both cases we retrieve the VMR of each gas as a single, altitude-independent scaling factor, and for the secondary eclipse we also retrieve the temperature as a function of pressure.

As found in Tinetti et al. (2010) and Barstow et al. (2013a), the radius assumed at the bottom of the model atmosphere has a significant effect on the transmission spectrum. This is because changes in the radius can move the whole spectrum up or down, and can also affect the atmospheric scale height; the scale height is inversely proportional to the gravitational acceleration, which in turn is inversely proportional to the square of the planet radius. Therefore, we now include the radius of the planet at 10 bar as a variable within the retrieval.

Barstow et al. (2013a) also find that, whilst there is not enough information to retrieve a continuous T-p profile in primary transit, the assumed temperature does affect the spectrum because the scale height is proportional to temperature. We therefore perform a bracketed retrieval of temperature, as in Barstow et al. (2013b), by performing the primary transit retrieval for a range of different temperatures and comparing the reduced-$\chi^2$ [1] values to find the best fit.

As we use only a single synthetic spectrum for each planet, we perform the secondary eclipse retrieval for three different temperature profile priors. This allows us to verify that the retrieval does not depend on the chosen prior, only on the measurement, provided that the spectrum contains information about the property to be retrieved.

### 2.1 Observing scenarios

The three observational scenarios for EChO (Varley et al. 2014) have different spectral resolving powers, and are designed to maximise the amount of information obtained from the greatest number of planets. For the most interesting, bright targets, the *Rosetta Stone* observing mode utilises the full spectral resolving power of the instrument, $R$=300 at wavelengths <5 μm and $R$=30 for λ>5 μm.

---

[1] $\chi^2/n$, where n = $n_{measurements}$-$n_{parameters}$ -1

This resolving power would be required to measure the abundances of minor trace species in the atmosphere and to constrain the vertical and temporal variation of gaseous species; this is used to characterise single targets in detail. The *Origins* mode works with $R$ for $\lambda<5$ μm degraded to 100, which results in a higher signal-to-noise ratio (SNR) at shorter wavelengths. This enables a larger number of planets to be surveyed in transmission and emission with the required signal to noise, and will allow the abundances of several gaseous species to be quantified. The *Chemical Census* observing mode operates at a spectral resolving power $R=50$ at wavelengths shorter than 5 μm, and will allow basic atmospheric characterisation of a large number of planets. In this scenario, a higher SNR can be obtained over fewer transits at the sacrifice of some spectral resolving power, allowing the temperature and most active molecular species to be constrained.

We perform retrieval tests for the different planet cases and observational scenarios, and check whether or not the recovered atmospheric state corresponds to the true state. If it does to within a reasonable margin of error (2 σ) then the true atmospheric state is theoretically recoverable from a similar EChO observation of a similar system.

## 3. Retrievals for model planets

We present the retrieval results for each of the model planet cases, including indications of the number of transits required in each case and which observational scenario is most favourable. As in Barstow et al. (2013a), we find that CO and $NH_3$ are generally poorly retrieved, probably due to their being fewer lines in the case of CO and a lack of high temperature lines in the list used for $NH_3$ (see Barstow et al. 2013a), so we do not show results for these gases.

### 3.1 Hot Jupiter

For all scenarios, assuming a single primary transit is observed, we recover all parameters correctly to within 2-σ for the lowest reduced $\chi^2$ case. This case corresponds to the retrieval

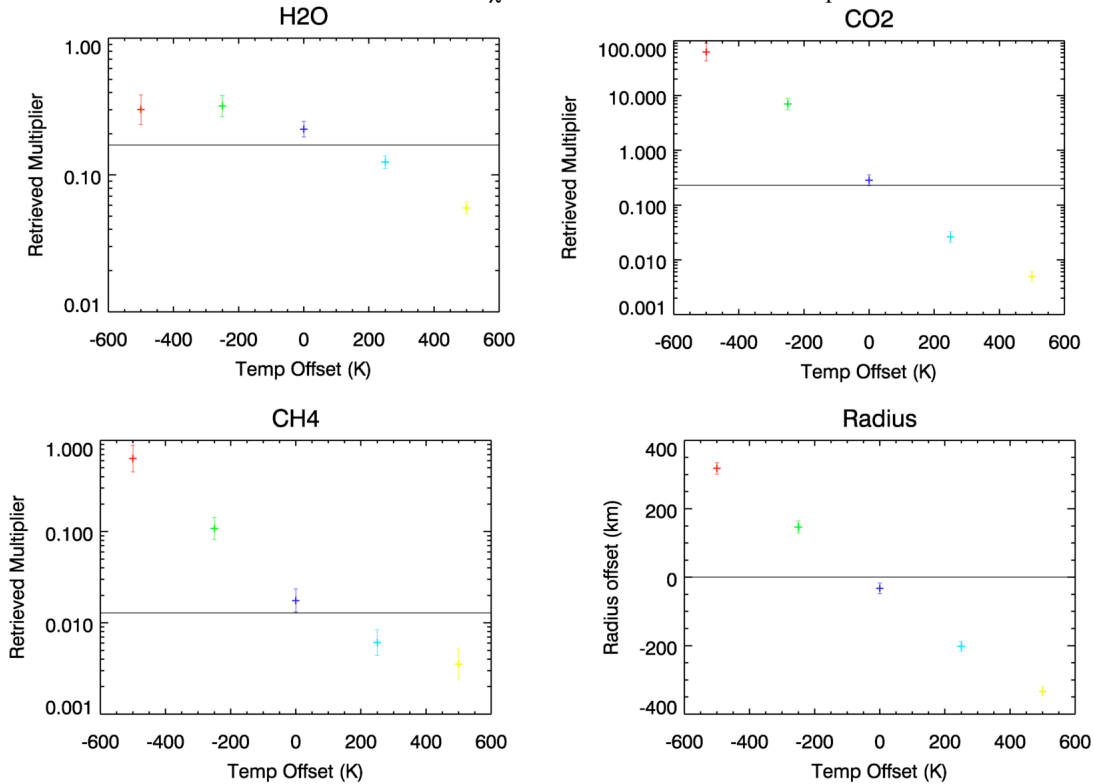

**Fig. 1.** Retrieved results for a single hot Jupiter primary transit in census mode. Temp offset indicates the offset from the input temperature profile in K. Solid horizontal lines indicate the input value of each parameter. Colours correspond to reduced $\chi^2$: red=10.4, green=1.8, blue=1.1 (best fit), turquoise=1.4, yellow=2.7

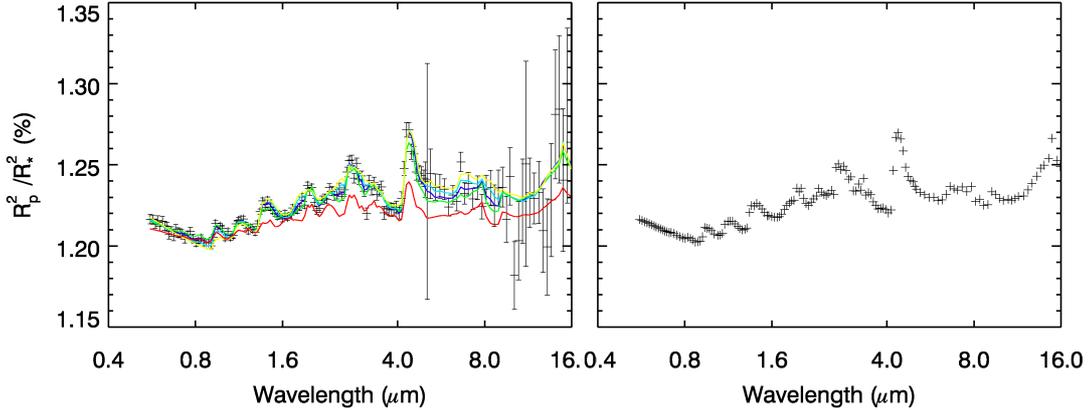

**Fig. 2.** Left – spectra for a single hot Jupiter primary transit in *Census* mode. The fitted spectra colours correspond to those in Figure 1. Right – the noise-free synthetic spectrum is shown for comparison

run using the input T-p profile, suggesting that we can also recover enough information about the temperature to make primary transit retrievals viable for cases where we cannot obtain a secondary eclipse observation, or the terminator temperature is very different from the dayside temperature. The *Chemical Census* retrieval results and spectra are shown in Figures 1 & 2. In Figure 2, for comparison, we also show the perfect input spectrum before noise was added. It is clear that the effect of the noise is increased at longer wavelengths.

Going from the *Census* case through to the *Rosetta Stone*, the difference in reduced $\chi^2$ between different temperature profile runs decreases; this is because the SNR of the spectrum decreases due to the higher resolution of the *Rosetta* spectrum, and so there is a wider range of models that can provide a fit with reduced $\chi^2$ <1.1 (Figure 3). In the *Census* case, only the correct T-p profile produced a good enough fit, whereas in the *Rosetta* case T-p +/- 250 K also produced an adequate fit; however, for a T-p profile shifted by 250 K the retrieved values are not correct due to complex degeneracies between temperature, radius and other parameters. To achieve similar reduced $\chi^2$ values for the *Rosetta* case as we find for the *Census*, we would need to observe 6 transits[2] as the resolution change is a factor of 6. This suggests that, for primary transit, higher resolution does not improve the retrieval of the altitude-independent abundance of the molecules considered here. A higher resolving power would potentially enable the retrieval of VMR as a function of altitude, but we do not consider this case here.

For the secondary transit, we assume a single eclipse is observed for all scenarios. As for the primary transit case, all gas VMRs are correctly recovered to within 2-sigma for all scenarios. However, we see differences in the retrieved T-p profile between the *Census*, *Origin* and *Rosetta* cases (Figure 4). It is clear that for the *Census* case the spectral resolution is too low to fully break the degeneracy between temperature and gas VMR, so the retrieved profile is less accurate. The improvement for the higher resolution cases is noticeable, demonstrating the importance of higher resolving powers for breaking degeneracies. Examples of spectral fits for the *Rosetta* case are shown in Figure 5 – the temperature prior chosen does not affect the retrieval or the spectral fit.

**3.2 Warm Jupiter**

We found in Barstow et al. (2013a) that a warm Jupiter orbiting a sun-like star at 35 pc would require ~30 transits/eclipses to ensure that the atmospheric state is correctly recovered. Viewing such a planet at lower spectral resolution may reduce the number required to obtain sufficient SNR.

We test the number of transits required to distinguish between the different temperature profiles for the *Rosetta* case. As for the hot Jupiter, in all cases the real input temperature profile produces the retrieved spectrum with the best-fit, but the reduced $\chi^2$ values for a single transit are all close to 1 and therefore we could not distinguish between the different retrievals and say which is correct. Retrieving an incorrect temperature results in the retrieved gas VMRs being incorrect too, due to solution degeneracy. Increasing the number of transits to 10 improves the situation to the extent that we could exclude the +/-150 K T-p profiles, but the +/-75 K fits are still acceptable. If we average over 20 transits, we could exclude all of the 'wrong' T-p profiles except -75 K. We show the noisy spectra and fits for these cases in Figure 6, and the retrieval results for the case for 20 transits in Figure 7.

---

[2] This is puzzling. In the photon noise limit the *Census* and *Rosetta* spectra should have the same information content, with the higher SNR for the *Census* observation offset by its lower *R*/fewer data. We would therefore expect the *Rosetta* case to be at least as constraining, if not more so, than the *Census* case. We only have a single test spectrum for each planet here, making it difficult to fully explore all effects of noise, and we have used the reduced $\chi^2$ as a simple discriminator. However, our results certainly suggest that for primary transit increasing *R* from 50 to 300 adds little information.

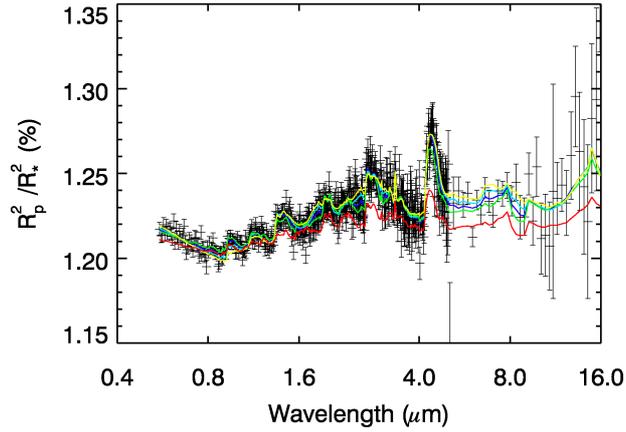

**Fig. 3.** Spectra for a single hot Jupiter primary transit in *Rosetta* mode, with different input T-p profiles. The fitted spectra colours correspond to different reduced $\chi^2$, (red: highest, black: lowest); reduced $\chi^2 \sim 1$ for the green, blue and turquoise synthetic spectra

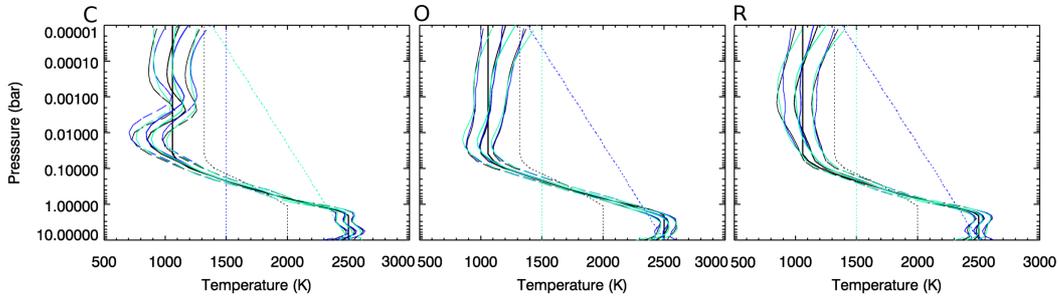

**Fig. 4.** Temperature retrievals from secondary transit observations (L-R: *Census, Origin, Rosetta* scenarios). The three temperature priors used are shown by dotted lines; the thick black line is the input profile, and the three retrieved profiles/error are shown by the thin solid/dashed lines. The model spectrum probes the atmosphere between $\sim 10^{-4}$ and $\sim 5$ bar

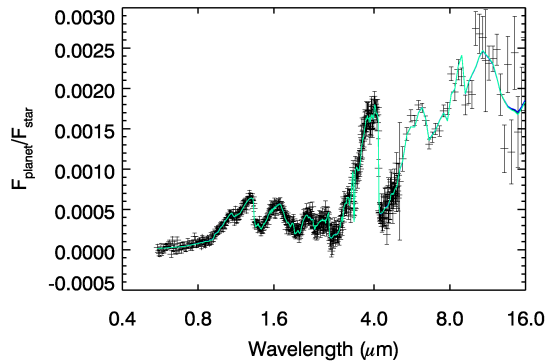

**Fig. 5.** Spectra for a single hot Jupiter secondary transit in *Rosetta* mode. The fitted spectra colours correspond to different temperature priors, as above. The temperature prior used does not affect the resultant spectral fit

The higher SNR for the *Census* and *Origin* observing scenarios should mean that 20 transits or fewer would be sufficient to recover the atmospheric state.

The warm Jupiter is much cooler than the hot Jupiter, and so there is significantly less thermal emission coming from the planet, and therefore the planet:star flux ratio in the secondary eclipse is much lower. This makes it harder to obtain information from a secondary transit for this planet, and multiple observations will be required. The number of observations depends on the observational scenario used. In Figure 8, we show retrieved temperatures and spectra for the *Census* and *Rosetta* cases, with 20 and 30 eclipses averaged over respectively. The quality of the spectral fit and the retrieved T-p profile is similar for these cases. As for the hot Jupiter, there is an advantage in higher resolution for breaking degeneracies, but a larger number of observations are required as increasing the resolution decreases the SNR. The stratospheric and deep temperatures are not well retrieved, leading also to poorer retrievals for the $H_2O$ and $CO_2$ abundances from the secondary eclipse than from the primary transit, due to solution degeneracy.

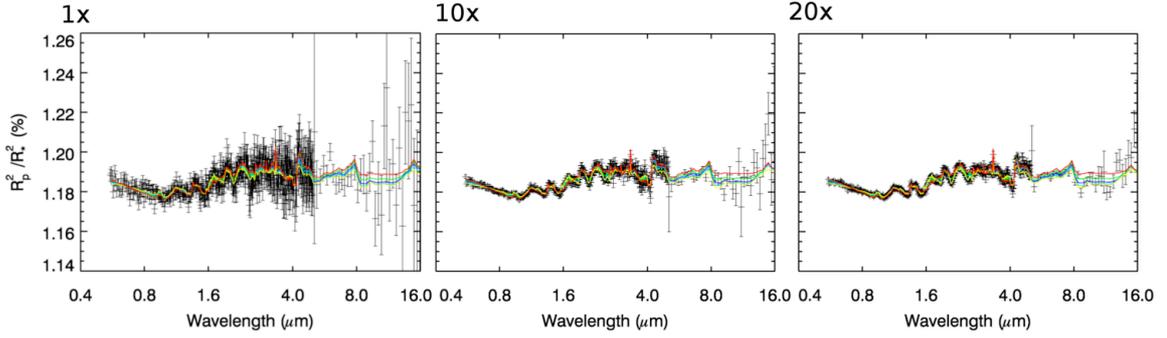

**Fig. 6.** Spectra for a warm Jupiter observed at *Rosetta* resolution, for 1, 10 and 20 transits. The colours of the fitted spectra correspond to reduced $\chi^2$ as before, with red the poorest fit followed by yellow, green, turquoise, blue. In the left-hand plot, all the spectra provide an adequate fit (reduced $\chi^2 \sim 1$). In the right-hand plot, only turquoise and blue provide an adequate fit. In all cases, the blue spectrum corresponds to the case where the T-p profile was the same as the input

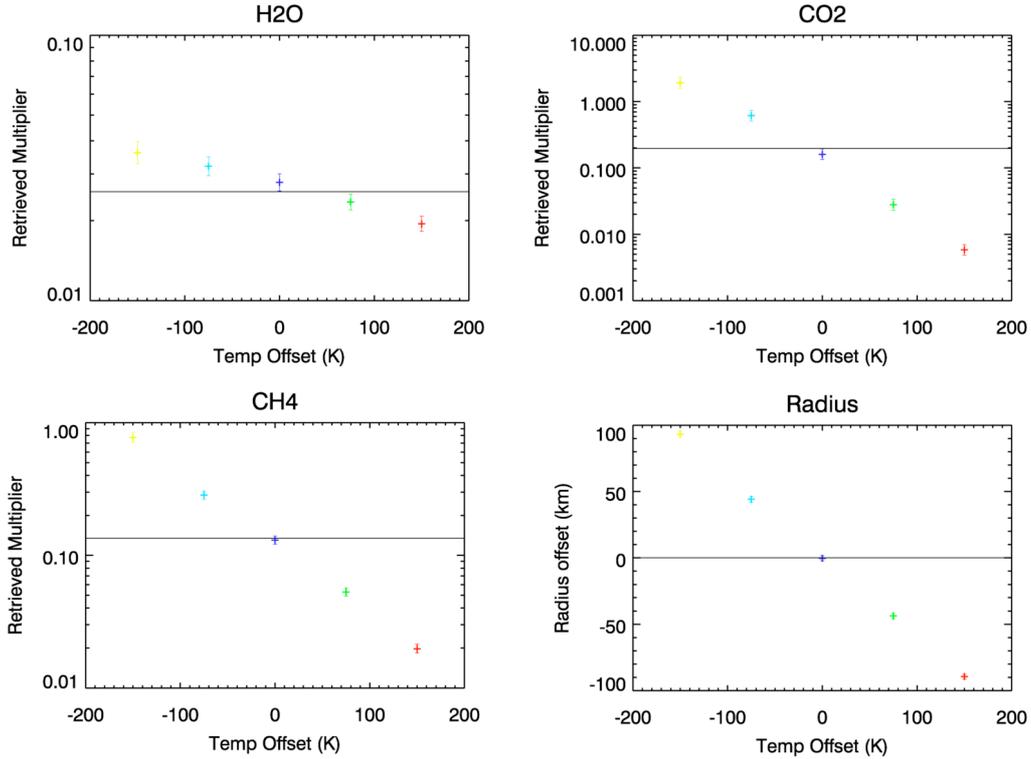

**Fig. 7.** Retrieved results for 20 warm Jupiter primary transits in *Rosetta* mode. Temp offset indicates the offset from the input temperature profile. Solid horizontal lines indicate the input value of each parameter. Colours correspond to reduced $\chi^2$: red=2.3, green=1.3, blue=1.0 (best fit), turquoise=1.1, yellow=1.7. For a good retrieval the reduced $\chi^2$ should be close to 1

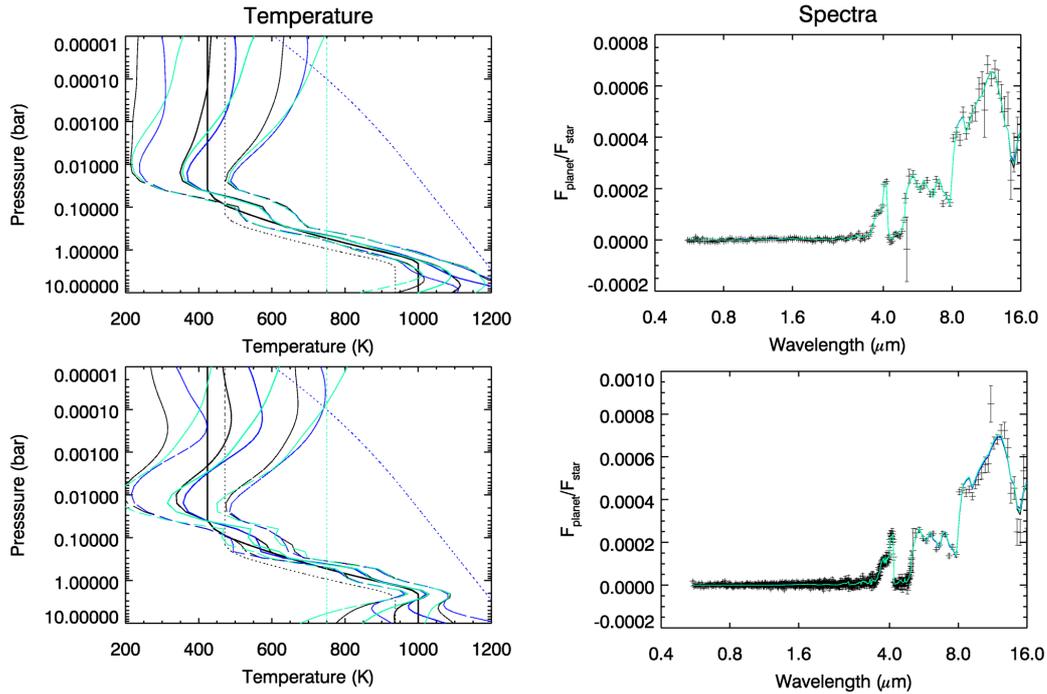

**Fig. 8.** Retrieved results for 20 warm Jupiter secondary transits in *Census* mode (top), and 30 in *Rosetta* mode (bottom). The spectral fits for each case are shown on the right. Colours correspond to the different temperature priors (shown by dotted lines). The thick, black solid line in the temperature plots shows the input temperature profile. The tropospheric temperature is reasonably well retrieved in both cases, but the temperature in the stratosphere and deep atmosphere cannot be reliably retrieved without further observations

### 3.3 Hot Neptune

The hot Neptune planet is in orbit around an M dwarf only 6 pc distant; the size and temperature of the star relative to the planet make this a favourable target for EChO in primary and secondary transit. Barstow et al. (2013a) found that a single transit or eclipse observation should be sufficient to recover the atmospheric state, but we now test this with the more realistic noise estimates provided by EChOSim.

We find that, for the *Rosetta* case in primary transit, we need 10 transits to be able to distinguish between the model fits with different T-p profiles. The results for 10x *Rosetta* transits are shown in Figure 10; spectra for 1x and 10x *Rosetta* transits are shown in Figure 9. This means that 10 transits or fewer would be required to distinguish between different T-p profiles for the *Census* and *Origin* modes.

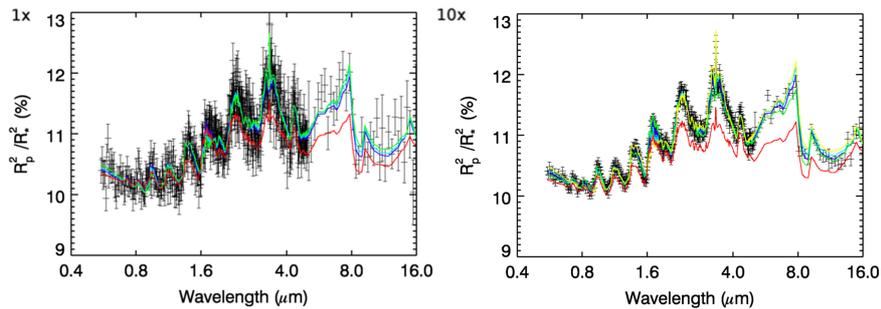

**Fig. 9.** Noisy spectra and fits for 1x and 10x hot Neptune transits in *Rosetta* mode. Reduced-$\chi^2$ is indicated by colour (red worst, blue best) for different T-p profiles. For 1 transit, green, turquoise and blue fits are indistinguishable in quality; for 10 only the blue model (correct T-p profile) fits well

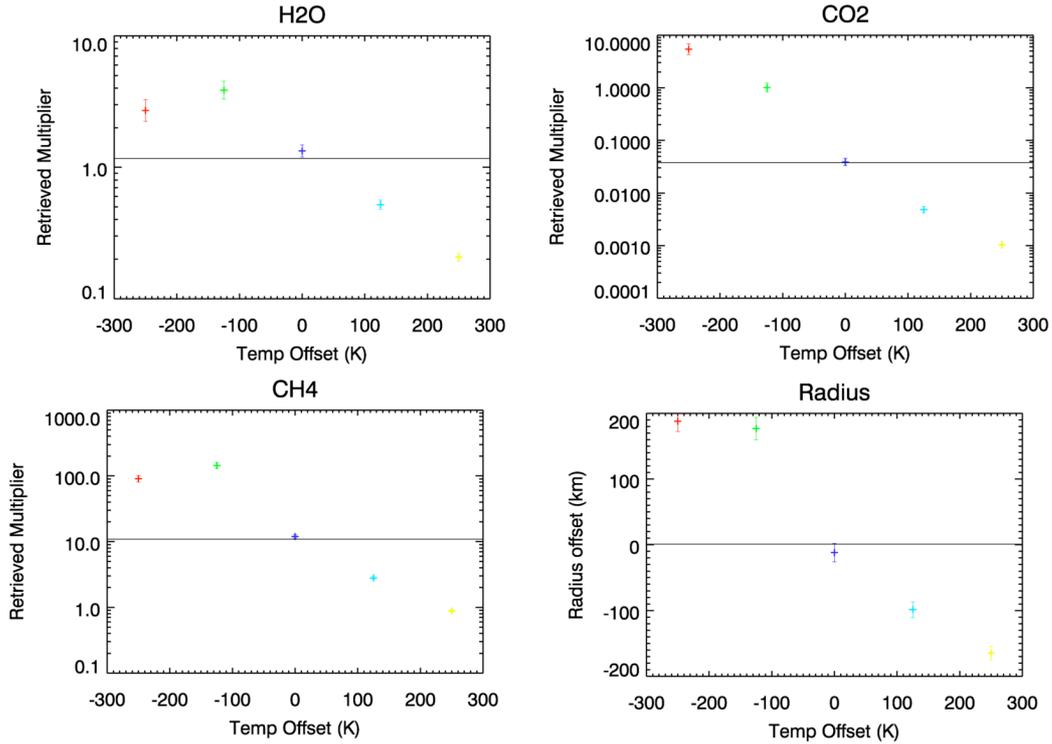

**Fig. 10.** Retrieved results for 10 hot Neptune primary transits in *Rosetta* mode, with different T-p profiles. Colours correspond to different reduced $\chi^2$ : red=17.1, green=2.0, blue=1.2 (best fit), turquoise=1.6, yellow=2.8. For a good retrieval the reduced $\chi^2$ should be close to 1

As for the primary transit, for the secondary transit we find here that we need multiple observations of the hot Neptune to accurately recover the atmospheric state. In Figures 11 – 12 we show noisy spectra and retrieval results for both of these cases. With 10 eclipses at *Rosetta* resolution the temperature and gaseous abundances are retrieved correctly. As this kind of planet is close to the parent star and therefore has a short period, obtaining multiple observations is not likely to be challenging. The hot Jupiter test demonstrated the importance of spectral resolution for breaking temperature/gas degeneracies, so we focus on the *Rosetta* mode here.

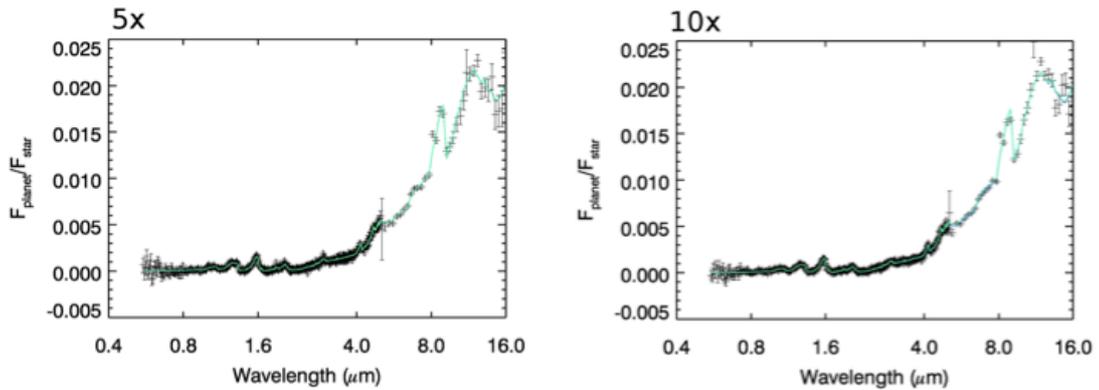

**Fig. 11.** Spectra for 5x hot Neptune eclipses (top) and 10x (bottom) in *Rosetta* mode. Colours of fitted spectra correspond to temperature priors as in Figure 12

## 4. Retrieval results for GJ 1214b

The questions of interest that we hope EChO will solve for GJ 1214b are somewhat different to those we have explored for the three synthetic planets. GJ 1214b is a warm super-Earth orbiting an M dwarf, for which we have compelling evidence from its transmission spectrum that its atmosphere contains some cloud or haze (Kreidberg et al. 2014). However, we have not yet obtained a secondary eclipse measurement for this planet since it is too cold, so its temperature structure is unknown; the bulk composition of its atmosphere is also unknown, as molecular absorption features in transmission are so far either obscured by cloud or not observed with sufficient spectral resolving power or SNR. EChO's unprecedented wavelength coverage would shed some light on this mysterious planet, by breaking degeneracies in retrievals of primary transit spectra and by enabling the detection of a secondary eclipse.

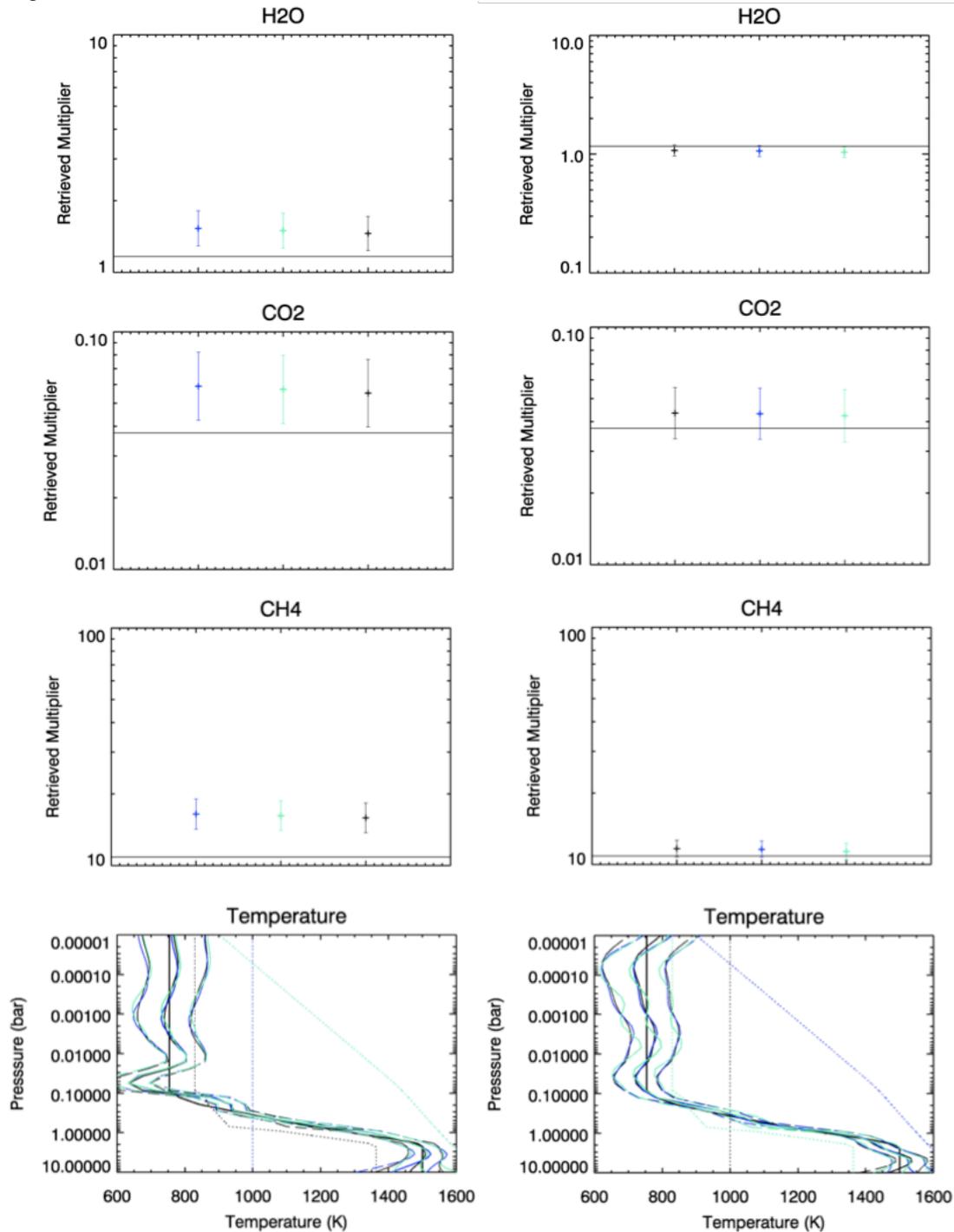

**Fig. 12.** Retrieved results for 5 (left) and 10 (right) hot Neptune eclipses in *Rosetta* mode. Colours correspond to different temperature priors (shown by dotted lines). The thick, black solid line shows the input temperature profile. Most gases are retrieved correctly to within 2-sigma for 5 eclipses, but the temperature at the tropopause is not retrieved well, due to degeneracy with the gas VMR retrievals. The temperature retrieval is much more accurate for 10 eclipses, and the gas VMR retrievals also improve

Kreidberg et al. (2014) show that a model cloud top pressure of 0.1-0.01 mbar or less is required to fit their *Hubble/Wide Field Camera 3* observations of GJ 1214b regardless of the bulk atmospheric composition assumed. A cloud top high in the atmosphere makes transmission spectra appear to be flat by making the atmosphere opaque up to high altitudes across a broad range of wavelengths. We test the effect of the cloud top pressure on our ability to retrieve gas volume mixing ratios for 30 *Census* transit observations of GJ 1214b.

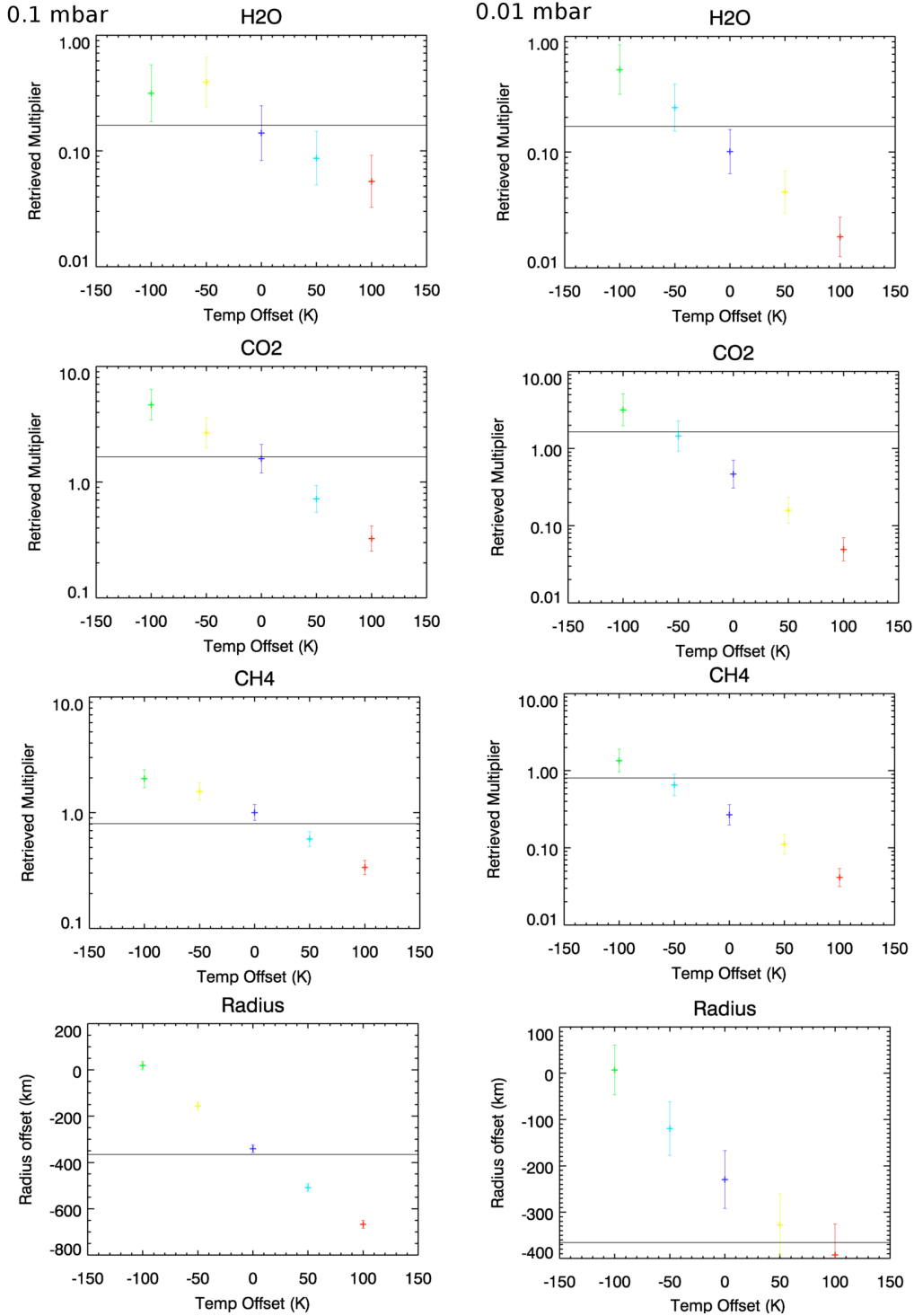

**Fig. 13.** Retrieved results for 30 transits of a simulated $H_2$-He-dominated GJ 1214b with a cloud top at 0.1 mbar (left) and 0.01 mbar (right). Colours/lines are as in Figures 1, 7 and 9. We obtain the correct temperature, radius and abundances if the cloud top is at 0.1 mbar but not if it is at 0.01 mbar, although even when the cloud is at 0.1 bar the reduced-$\chi^2$ values are similar for all retrievals

We present primary transit bracketed retrieval results for cloud top pressures of 0.1 and 0.01 mbar. Whilst we cannot accurately retrieve the cloud optical depth, we can retrieve the gas VMR and 10-bar radius correctly if the cloud top is at 0.1 mbar, but not if it is at 0.01 mbar (Figure 13). A higher cloud top makes it significantly more difficult to retrieve gas VMRs, as the absorption features are masked to a greater extent. However, it has been demonstrated by Kreidberg et al. (2014) that the cloud top pressure can be constrained, so if the cloud top pressure is found to be low we would know that further transits would be required in order to constrain the atmospheric properties of GJ 1214b. This is an important illustration of the effect of clouds on transmission spectra, and the challenges of transit spectroscopy for smaller planets.

We cannot fit the synthetic cloudy $H_2$-He GJ 1214b spectrum with an atmospheric model with increased water vapour abundance and a higher molecular weight, as found by Barstow et al. (2013b), indicating that we could distinguish between atmospheres of different molecular weights for GJ 1214b.

With EChO, we could also place constraints on the stratospheric temperature from observations of the secondary eclipse. This has not yet been achieved as the temperature contrast of GJ 1214b with respect to its host star is low, but we show here that it would be possible with EChO. In Figure 14, we show that with 30 secondary eclipses and rebinning to R=10 for wavelengths longer than 5 microns, we can obtain some spectral information from a secondary eclipse of GJ 1214b. Because GJ 1214b is cool there is no information in the eclipse spectrum for wavelengths shorter than 5 μm, emphasising the importance of a large spectral range for this kind of observatory. The signal to noise is insufficient for us to place constraints on the composition of the dayside atmosphere, but we demonstrate in Figure 15 that we can retrieve the temperature from 100–1 mbar for an $H_2$-He-dominated atmosphere and 100–10 mbar for an $H_2O$-dominated atmosphere. This would significantly enhance our ability to constrain the terminator properties from the primary transit, by providing a first measurement of temperature.

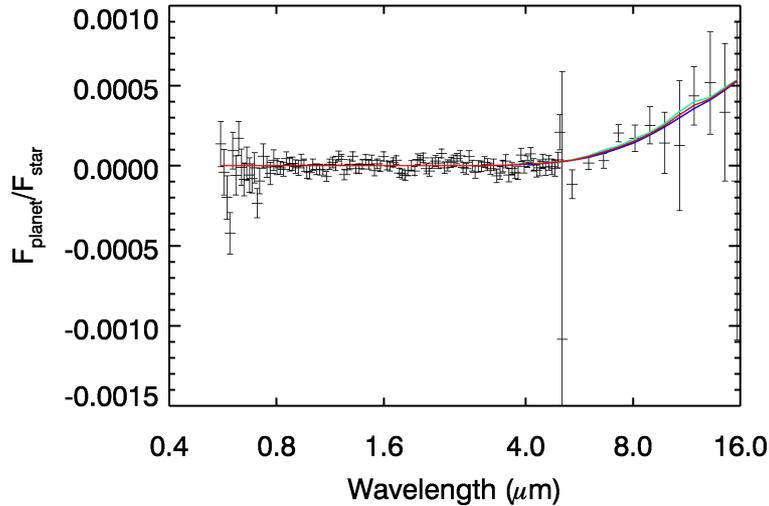

**Fig. 14.** Spectrum for 30 secondary eclipses of GJ 1214b, with fluxes rebinned to R=10 longwards of 5 microns. This is for the $H_2$-He dominated case, but the $H_2O$ dominated case is similar. Model fits for different temperature priors are shown by coloured lines (all fit equally well)

## 5. Discussion and Conclusions

Here, we have demonstrated the efficacy of EChO for characterising a range of planets with $H_2$-He dominated atmospheres. For hot Jupiters and Neptunes, retrieval of gas volume mixing ratios, temperature structure and radius is possible with only a handful of transmission and eclipse spectra. Over the lifetime of the mission, this translates to being able to observe potentially hundreds of objects, providing an unprecedented sample of extrasolar hot atmospheres to begin to understand what drives the differences between these worlds.

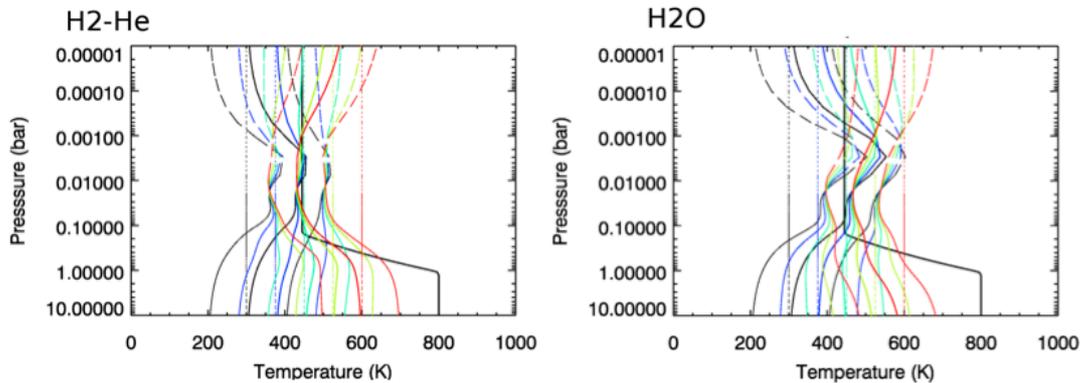

**Fig. 15.** Temperature retrievals for 30 secondary eclipses of GJ 1214b, assuming an $H_2$-He dominated composition (left) and an $H_2O$-dominated composition (right). The different priors are shown by dotted lines, the retrieved temperatures by solid lines and the error by dashed lines. The thick black line is the input profile. The retrievals for both atmospheric scenarios converge at a pressure level of around 20 mbar, indicating that we can retrieve the temperature at this pressure level

EChO also represents a significant advance on our current capability to observe super-Earth atmospheres in transit. Its large wavelength coverage, especially longwards of 5 μm, will enable secondary eclipses of cooler objects to be measured for the first time, and will also provide the first real prospect of breaking degeneracies between cloud and other atmospheric properties.

We find that we would require only a single transit of a hot Jupiter around a sun-like star at 35 pc to retrieve gaseous abundances and radius from the terminator, and gas abundances and temperature from the dayside, for the *Census* and *Origins* observing modes. To achieve sufficient SNR for the *Rosetta* mode we would require 6 transits, but only one eclipse. For a warm Jupiter orbiting a similar star, we would require 20 transits or eclipses, with 30 eclipses required to correctly retrieve the dayside temperature profile (at altitudes where the spectrum contains information) for a *Rosetta* mode observation. For the case of a hot Neptune orbiting an M dwarf at 6 pc, we would require 10 transits and eclipses for all cases. For these targets, obtaining more observations to achieve sufficient SNR in the *Rosetta* observing mode would potentially allow rarer trace species than those modelled here to be characterised. In addition, the greater wavelength coverage of EChO allows us to probe a larger altitude range than any previous instrument, and therefore to more fully constrain the T-p profile.

We find that, depending on the cloud properties of GJ 1214b, we would require at least 30 transits and possibly more to constrain the composition of the terminator atmosphere. GJ 1214b is too cool to retrieve the atmospheric composition on the day side, but with 30 eclipses it would be possible to measure the stratospheric temperature of the planet. This would greatly enhance our level of knowledge about GJ 1214b, and we expect that these findings would also apply to other, similar super-Earths.

## 6. Acknowledgements

JKB is supported by the Science and Technology Facilities Council and LNF by a Royal Society Research Fellowship. This work was part of the EChO Consortium Study funded by the UK Space Agency. We thank the anonymous reviewer and Professor Jacob Bean for helpful comments on the manuscript.